# Superconducting memory and trapped magnetic flux in ternary lanthanum polyhydrides

Dmitrii V. Semenok[1,†,*], Andrey V. Sadakov[2,†], Di Zhou[1,*], Oleg A. Sobolevskiy[2], Sven Luther[3], Toni Helm[3], Vladimir M. Pudalov[2,4], Ivan A. Troyan[5,†] and Viktor V. Struzhkin[1,6,*]

[1] *Center for High Pressure Science & Technology Advanced Research, Bldg. #8E, ZPark, 10 Xibeiwang East Rd, Haidian District, Beijing, 100193, China*

[2] *V. L. Ginzburg Center for High-Temperature Superconductivity and Quantum Materials, Moscow, 119991 Russia*

[3] *Hochfeld-Magnetlabor Dresden (HLD-EMFL) and Würzburg-Dresden Cluster of Excellence, Helmholtz-Zentrum Dresden-Rossendorf (HZDR), Dresden 01328, Germany*

[4] *National Research University Higher School of Economics, Moscow 101000, Russia*

[5] *A.V. Shubnikov Institute of Crystallography of the Kurchatov Complex of Crystallography and Photonics, 59 Leninsky Prospekt, Moscow 119333, Russia*

[6] *Shanghai Key Laboratory of Material Frontiers Research in Extreme Environments (MFree), Shanghai Advanced Research in Physical Sciences (SHARPS), Pudong, Shanghai 201203, China*

*Corresponding authors: Di Zhou (di.zhou@hpstar.ac.cn), Viktor Struzhkin (viktor.struzhkin@hpstar.ac.cn), and Dmitrii Semenok (dmitrii.semenok@hpstar.ac.cn)

†These authors contributed equally.

## Abstract

Superconducting memory is a promising technology for data storage because of its speed, high energy efficiency, non-volatility, and compatibility with quantum computing devices. However, the need for cryogenic temperatures makes superconducting memory an extremely expensive and specialized device. Ternary lanthanum polyhydrides, due to their high critical temperatures of 240-250 K, represent a convenient platform for studying effects associated with superconductivity in disordered granular systems. In this work, we investigate a trapped magnetic flux and memory effects in recently discovered lanthanum-neodymium $(La,Nd)H_{10}$ and lanthanum-scandium $(La,Sc)H_{12}$ superhydrides at a pressure of 175-196 GPa. We use a steady magnetic field of a few Tesla (T) and strong pulsed fields up to 68 T to create the trapped flux state in the compressed superhydrides. We find a clockwise hysteresis of magnetoresistance in cerium $CeH_{9-10}$ and lanthanum-cerium $(La,Ce)H_{10+x}$ polyhydrides, a characteristic feature of granular superconductors. A study of the current-voltage characteristics and voltage-temperature curves of the samples with trapped magnetic flux indicates a significant memory effect in La-Sc polyhydrides already at 225-230 K.

**Graphical abstract**

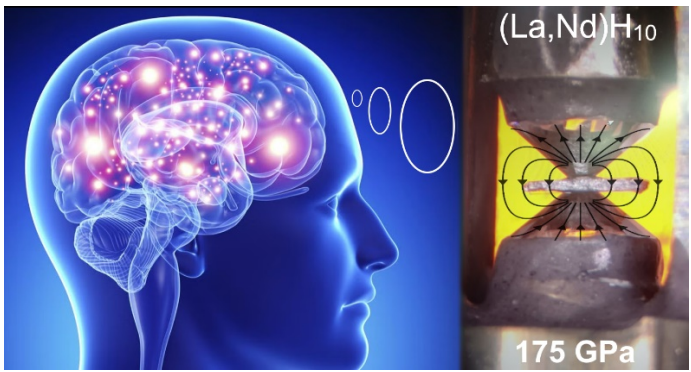

# Introduction

The effects associated with "freezing" of magnetic flux in granular superconductors, for instance, cuprates, are fairly well known [1,2]. Such samples will exhibit a remnant magnetization, indicating flux trapping, when they are cooled in an external magnetic field and then the magnetic field is removed. This effect induces hysteresis in various magnetotransport properties, in particular magnetoresistance [3,4] and critical current [1,2,5]. The heterogeneity of polyhydride samples at high pressure leads to the appearance of various physical effects: diode conductivity [6], SQUID (superconducting quantum interference device) effect [7], complex dynamics of the vortex lattice [8,9], and reduced flux expulsion under field cooling [10]. The latter may happen since the magnetic field remains trapped both in intergranular regions and by pinning centers inside micro grains.

Freezing of magnetic flux in superconductors is usually studied using SQUID magnetometers [11] and microscopes [12]. This phenomenon enables electronic memory devices, where storage of information is based on magnetic flux vortices frozen in nanoscale superconductor granules, for example, in Co/Nb heterostructures [13]. However, the use of SQUID magnetometers for systematic a study of samples at high pressure is complicated because of extremely small sample sizes and the need to subtract significant background signal originating from high-pressure diamond anvil cells (DACs) [11]. In this paper, we demonstrate a simpler method of detecting trapped magnetic flux. We observe the appearance of a hysteresis in the current-voltage (*V-I*) characteristics and a change (decrease) in the offset critical temperature of the superconducting transition (Figure 1b). The existence of this hysteresis was proposed as one of the fingerprints for superconductivity in polyhydrides by J. Hirsch in 2022 [14] and is studied in this work. This storing of flux can be described in terms of a writing and erasing a memory.

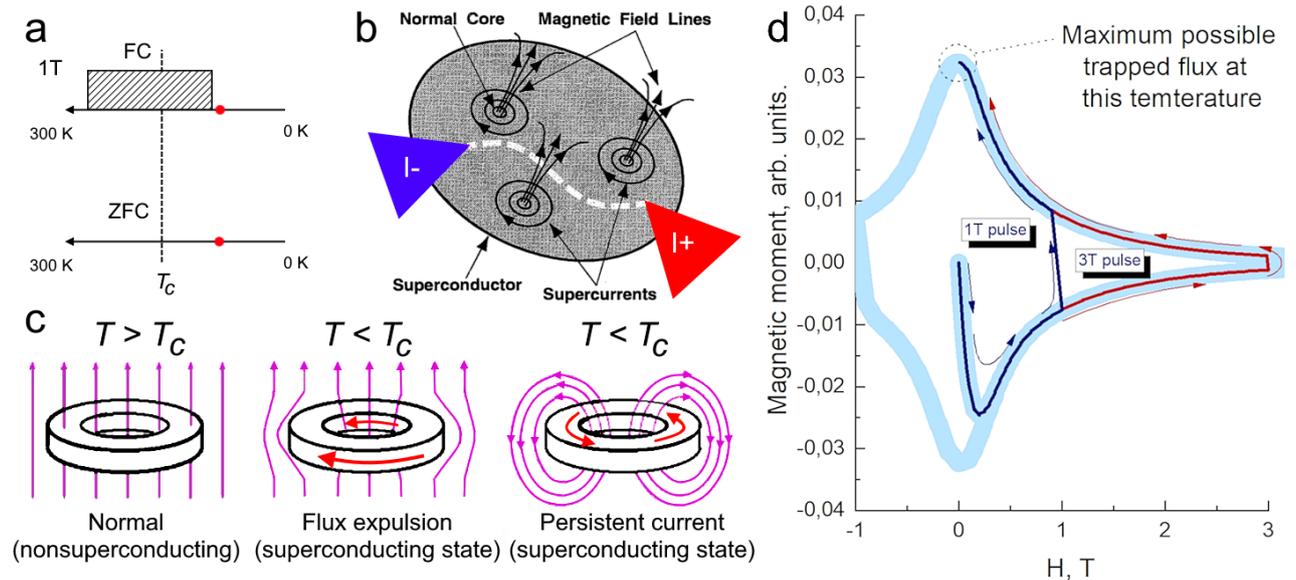

**Figure 1.** Schematic diagram of the experiment with magnetic flux trapping and its effect on the critical current in "weak" links of the SC samples. (a) Schematic diagram of the experiment with comparison of *V-I* or *V-T* characteristics recorded in the zero-field cooling (ZFC) mode and in the field cooling (FC, 1T) mode. The DC field is switched off and set to zero below the critical temperature ($T_C$). The red dot is the temperature at which the measurements are taken. (b) Sketch of the electric current flowing through the sample with trapped vortices. "I+" and "I−" correspond to the current electrodes. (c) Schematic diagram of the emergence of trapped magnetic flux when the field penetrating through the superconducting ring is switched off. The field induced shielding current is indicated by red arrows, and the magnetic field flux lines shown by purple arrows. (d) A typical hysteresis loop of a type II superconductor at a certain temperature $T < T_C$ (thick light blue line).

A superhydride sample under pressure in a DAC can be simplified as a logic element with only two states <0> (with trapped flux) and <1> (no trapped flux). In our work, the "memory erase" operation, i.e., changing <1> → into <0> is realized by a rapid warming of the sample above its superconducting critical temperature $T_C$(onset) into the normal conducting phase. The "write" operation is performed by changing the magnetic field induction when the sample is in the SC state (Figures 1a, c). For example, when introducing a magnetic field after the degauss and ZFC procedures, the magnetic moment of the superconductor goes from zero, surpasses the lower critical field $B_{C1}$, reaches the hysteresis loop following the dark blue arrow (up to 1 Tesla, Figure 1d) or the red arrow (up to 3 Tesla). When the desired target field is reached, the field decrease begins. The magnetic moment moves to the upper branch of the hysteresis loop and reaches its maximum at $B = 0$. If the target field is much greater than the lower critical field (and in our case this condition is met), then for each specific temperature the maximum value of the captured flux will be the same.

In this work, we show that slow controlled cooling and heating of La-Nd and La-Sc superhydrides lead to almost completely coinciding and reproducible *V-I* and *V-T* (voltage-temperature) dependencies. By contrast, the incorporation of a magnetic field pulse into the sample cooling-warming protocol induces a hysteresis, a decrease in the critical current ($I_C$) and an offset in the critical temperature of the sample $\Delta T_C$ from 0.1-1.5 K up to 3-4 K. This hysteresis completely disappears after the "erasure" of memory.

## Results and Discussion

*1. Trapped magnetic flux in (La, Nd)$H_{10}$*

The studied LaH$_{10}$ sample doped with 6.5 at. % of Nd, synthesized from corresponding La-Nd alloy, placed in a NiCrAl diamond anvil cell at a pressure of 175 GPa. Since the sample had already been subjected to multiple heating and cooling cycles (it was prepared in March 2022 [15]), the transition to the superconducting state degraded from initial ≈ 180 K to $T_C \approx$ 162-174 K (Figure 2a). Nevertheless, when temperature decreased below 154 K, the sample still exhibits a "zero" resistance state (i.e., the noise of resistance measurements is higher than averaged resistance value, as shown in the inset to Figure 2a). In the first stage of the study of the (La,Nd)H$_{10}$ sample, we measured the current-voltage characteristics at temperatures from 150 K to 160.5 K in zero magnetic field. The obtained *V-I* curves are typical for superconductors: the voltage drop on the sample remains zero until the critical current value is reached (Figure 2c). The pronounced nonlinearity of the *V-I* curves, noted earlier for Ce superhydrides by E. Talantsev [16], is noteworthy. It is also interesting to note the absence of any pronounced features in the temperature dependence of the resistance *R(T)* in the range from 174 to 345 K. Thus, there is no evidence of additional high-$T_C$ resistive transitions (that would signify the presence of incidental phases), or high-temperature hydrogen diffusion in the studied sample [17].

At the second stage of the experiment, we created trapped magnetic vortices in the sample cooled in zero magnetic field from 190 K to 150 K. For this, we slowly (during 10 minutes) ramped magnetic field from zero to $B_{max}$ = 1 T and than back to zero followed by a degaussing procedure (see Figure 2d, inset). For subsequent studies of the *V-I* characteristics, four temperature points were selected: 150 K, 151 K, 152 K and 153 K. In all cases, we observed a noticeable hysteresis of the critical currents. For example, after turning off the magnetic field of 1 T, measurements of the *V-I* characteristic of the (La, Nd)H$_{10}$ sample at 150 K were carried out (Figure 2d). We found that the critical current value differs from the initial one measured in the zero-field cooling mode (ZFC, marked as "0"). The critical current became lower: $I_C$(<0>) = 8.4 mA → $I_C$(<1>) = 7.6 mA, $\Delta I_C$ = − 0.8 mA, that signifies to the presence of a memory about the passed triangular wave of the magnetic field. A simple estimate based on the dependence of the upper critical field on temperature $B_{C2}(T)$ shows that the trapped magnetic field can reach 90-240 mT. Indeed, the difference $\Delta I_C$ = –0.8 mA corresponds to the change $\Delta T_C$ = − 0.24 K (Figure 2c), that with a typical for hydrides derivative $dB_{C2}/dT|_{T=Tc} \approx -1$ T/K [15] gives a trapped field around 240 mT. We stress that during these measurements the sample temperature was controlled with accuracy of 0.05 K and hence, the temparture variation could not cause such $T_C$ change.

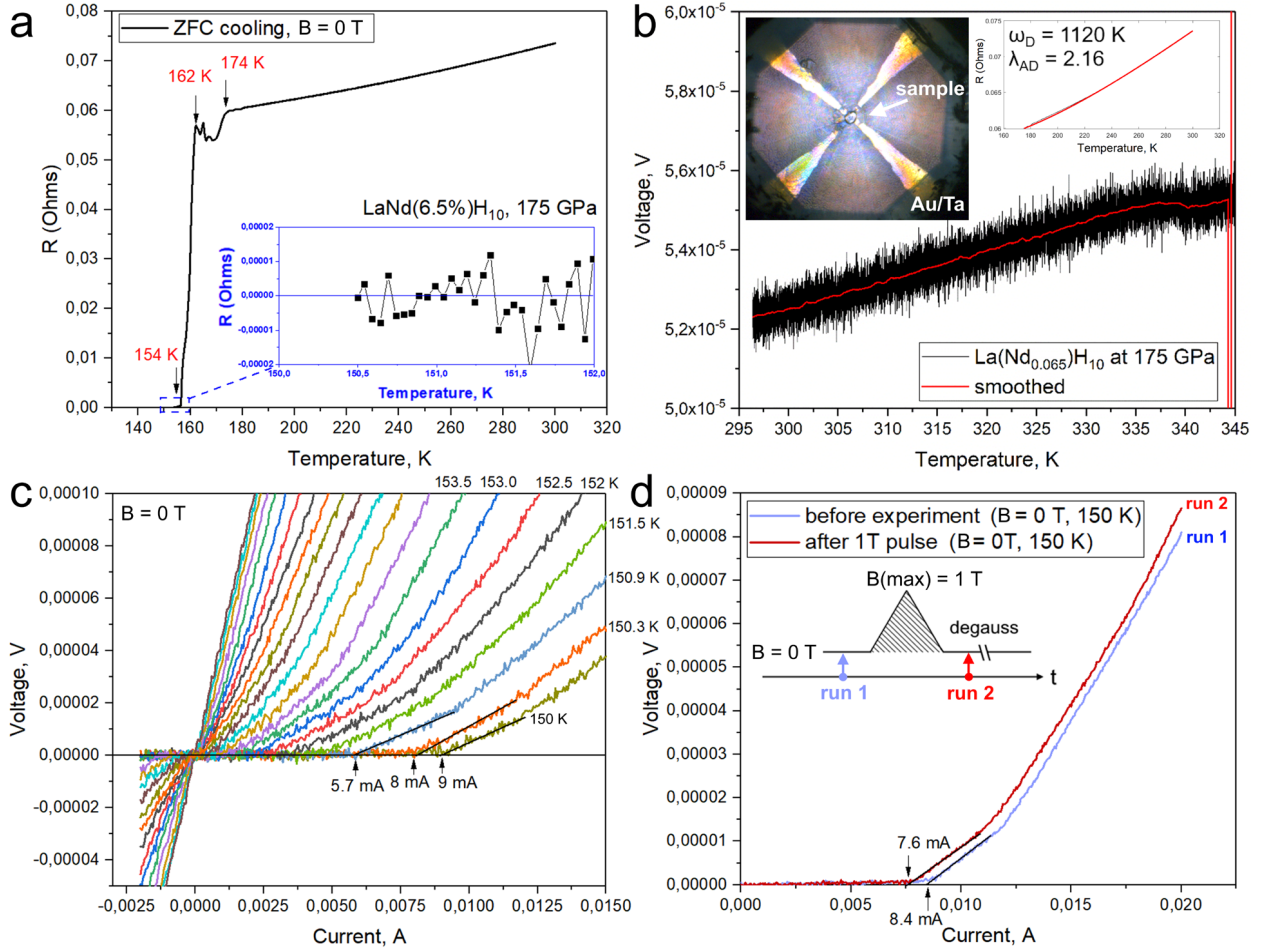

**Figure 2.** Transport properties of the (La,Nd)H$_{10}$ sample at 175 GPa. (a) Temperature dependence of the electrical resistance of (La,Nd)H$_{10}$ measured in a warming cycle with a DC current of 0.1 mA in the absence of a magnetic field. Inset: residual resistance of the sample below 152 K. (b) High-temperature part of the sample *V-T* dependence, which terminates at 345 K due to the destruction of the DAC. In general, the sample resistance increases linearly in the range 300-345 K. Inset 1: Photograph of the DAC's chamber with electrodes (Au/Ta) and hydride sample. Inset 2: Fit of the normal part of the electrical resistance using the Bloch-Grüneisen [18,19] and Allen-Dynes [20] formulas. Electron-phonon coefficient $\lambda_{AD}$ is calculated within the assumption $\mu^*=0.1$. (c) Current-voltage characteristics of the (La,Nd)H$_{10}$ at 175 GPa measured in the absence of a magnetic field at different temperatures from 150 K to 160.5 K in steps of 0.5-0.6 K. In this experiment, the current increased to a large "positive" value ($dI/dt > 0$). (d) Current-voltage characteristic of the sample at 150 K before (blue curve, run 1) and after (red curve, run 2) a triangular magnetic field pulse with a maximum amplitude of 1 T.

In the third part of the experiment with (La,Nd)H$_{10}$, we investigated *V-I* characteristics in a constant magnetic field of 1 T and 3 T in order to clarify the reversibility of the *V-I* measurements upon increasing and decreasing current ($dI/dt > 0$ and $dI/dt < 0$, indicated by arrows, Figures 3a,b). It was found that the *V-I* curves are well reproducible with a fairly good accuracy (~ 0.02 mA) both in the constant magnetic field (1 T and 3 T) and in the absence of external field. This reproducibility shows that the detectable trapped magnetic flux cannot be formed by passing an applied instrument current (0.1 mA) through the sample in most cases. The exception may be the case shown in Figure 3a (light blue curve), where a small hysteresis in *V-I* curves for ramping the current at steady field *B* = 1 T can still be observed. At the same time, the triangle magnetic field wave (with $B_{max}$ = 1 T) transfers part of the granular sample to the normal state. Then, when the magnetic field decreases, the sample goes into a superconducting state, capturing part of the magnetic flux, which can be detected by the shift in the *V-I* curves (Figure 3b, the hysteresis zone is highlighted in yellow).

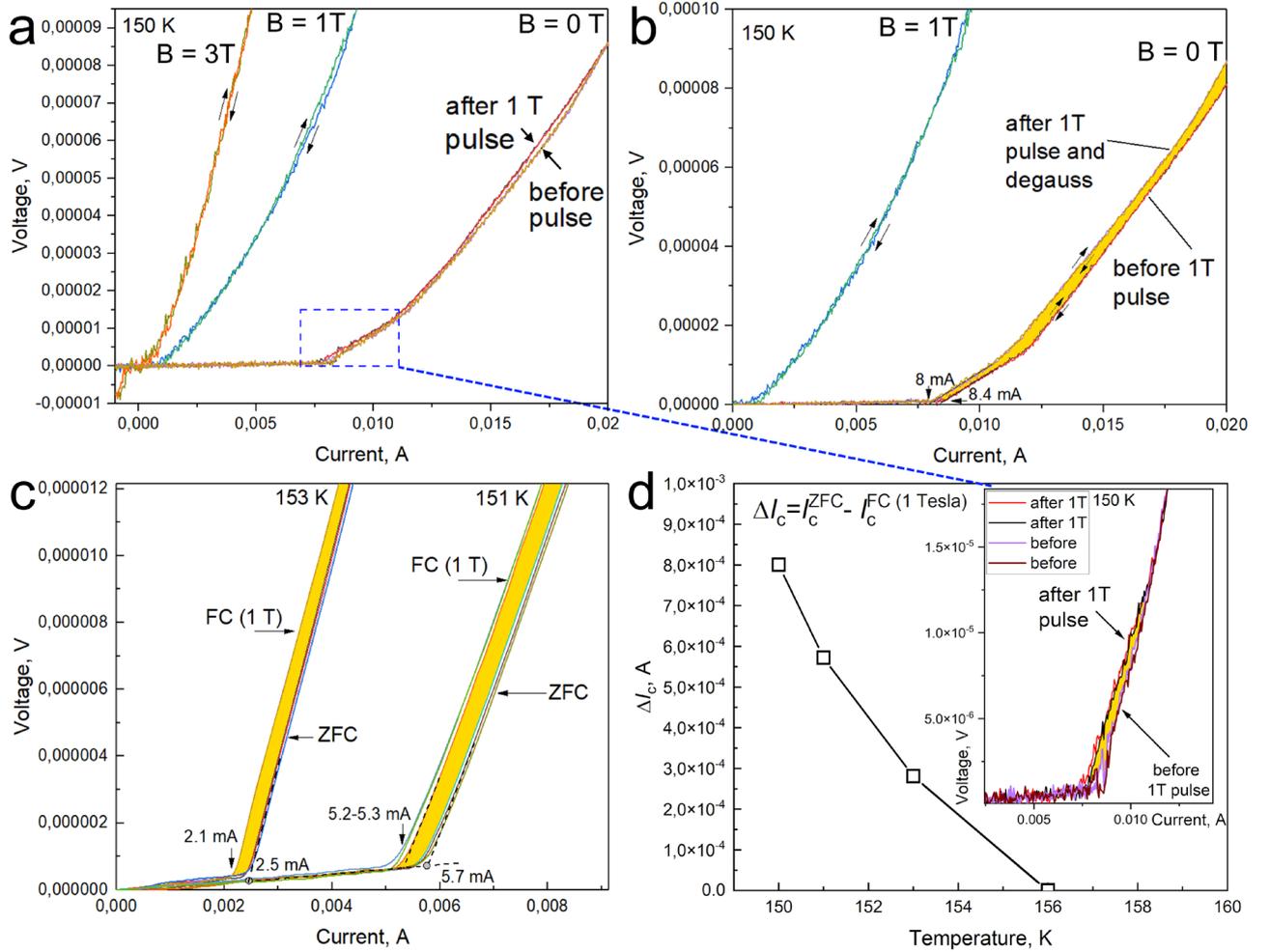

**Figure 3.** Current-voltage characteristics (*V-I*) of the (La,Nd)$H_{10}$ sample at 175 GPa, measured before and after magnetic flux capture. The hysteresis region is highlighted in yellow. (a) Current-voltage (*V-I*) curves of the sample at 150 K in constant magnetic fields (1 and 3 T) in the current increasing (*dI/dt* > 0, arrows up) and decreasing (*dI/dt* < 0, arrows down) modes compared to the *V-I* curves obtained in zero field with and without a trapped magnetic field. The corresponding hysteresis is shown in the inset of panel (d). (b) Current-voltage (*V-I*) characteristics of the sample at 150 K in a magnetic field of 1 T compared to *V-I* curves obtained in zero field with and without a trapped magnetic field ("triangle" *B(t)* plot) in the mode of increasing (*dI/dt* > 0, arrow up) and decreasing current (*dI/dt* < 0, arrow down). The corresponding hysteresis is highlighted in yellow. (c) Current-voltage characteristics (*V-I*) of the sample at temperatures of 151 K and 153 K, obtained during zero-field cooling (ZFC) and during cooling in a magnetic field of 1 T (FC) followed by its abrupt switching off. The experiment was performed three times and demonstrates good reproducibility. The corresponding hysteresis of the *V-I* curves is highlighted in yellow. (d) Dependence of the critical current difference $\Delta I_C = I_C(<0>) – I_C(<1>)$ on the field-switching temperature. Qualitatively, an increase in $\Delta I_C$ indicates a higher value of the trapped magnetic flux and a higher value of circulating eddy currents in individual SC granules.

A significantly larger magnetic flux can be captured within the classical protocol: ZFC → *V-I* measurements (*<0>* state) → warm up to $T > T_C$ → switch on the magnetic field → FC below $T_C$ → switch off the magnetic field → *V-I* measurements (*<1>* state). In this procedure (Figure 3c), after measuring the *V-I* characteristic in the ZFC mode and subsequent heating to $T > T_C$, we cool the sample below the critical temperature in a constant magnetic field, then switch the field off and compare the measurement results with the *V-I* of the sample cooled in the zero field. The sample memory is "erased" by heating it above 190 K.

To study the reproducibility of the effect, the experiment according to this protocol was repeated three times (see Figure 3c, triple curves). As a result, we found that there is a systematic deviation of *V-I* and the critical current $I_C$ value in the experiment with field cooling from the results obtained within the zero-field cooling (ZFC). At the same time, the difference $\Delta I_C = I_C(ZFC) - I_C(FC)$ increases with decreasing field turn-off temperature (Figure 3d), indicating that the ability to trap a magnetic flux increases with decreasing temperature. This is consistent with what is expected for type II superconductors, i.e., that the critical current and the amplitude of the magnetization hysteresis loop corresponding to the trapped magnetic flux increase with cooling.

*2. Memory effect in (La,Sc)H$_{12}$*

Experiments in strong pulsed magnetic field [6,15,21-23] at low temperatures are accompanied by random magnetic flux trapping in the granules of hydride superconductors. These trapped vortices can be the cause for a hysteretic temperature dependence of electrical resistance *R(T)* or voltage drop *V(T)*. Fore a proper detection, the current should remain constant in the same cooling/heating cycle with a controlled low sweep rate (0.5–1 K/min in our case). The temperature sensor and sample need to be well thermalized. It is important that the thermometer in such measurements should be tightly attached to the gasket and located as close as possible to the hydride sample.

The second sample, investigated in October-November 2023 at HZDR (Dresden, Germany), was novel lanthanum scandium hydride (La,Sc)H$_{12}$ at a pressure of 196 GPa [6]. This sample shows a high reproducible $T_C$(onset) ≈ 246 K. The superconducting transition in the sample ends at about 210-215 K (Figure 4d, f). In this experiment (Figure 4), we cooled the sample in zero magnetic field, stabilized a certain temperature to perform a 68 T pulse with 150 ms pulse duration. Thereafter, we warmed the sample with a constant rate of 0.5-1 K/min. At the same time, we measured the voltage on the sample (real part, Re) using an AC excitation current of 1 mA (RMS) with a frequency of 100 Hz. Since even at a slow temperature sweep rate of 0.5 K/min there is a hysteresis between the cooling and warming curves (e.g., Figures 4e, 5), we compare only cycles with the same *dT/dt* sign. Figures 4e-f show the corresponding hysterisises (yellow) that occur between the *V-T* curves of different magnetic pulse temperatures. At $T_{pulse}$ = 210 K there is almost no hysteresis below 225 K (Figure 4d). In this temperature range, the memory effect is below the experimental resolution limit (~0.02 K).

When the magnetic pulse is carried out at 201 K, there is a more pronounced hysteresis of the critical temperature $T_C$(offset) equal to $\Delta T_C ≈ 0.3$ K (Figure 4e). For pulses at 180 K and 193 K, the hysteresis continues to increase and reaches values of up to $\Delta T_C ≈ 1.7$ K in different regions of the *V(T)* curve (Figure 4f). It is important to note that the most pronounced memory effect in (La,Sc)H$_{12}$ is observed in the same temperature range where the diode and SQUID effects occur [6]. It is also logical that the transition to the normal state begins earlier (at 214.9 K) in the experiment where the "writing" into the sample memory was carried out at a lower temperature (180 K), and where the trapped magnetic flux and critical current are higher than at 193 K (Figure 4g). The study of lower temperatures from 185 K to 172 K indicates the same trend (Figure 5).

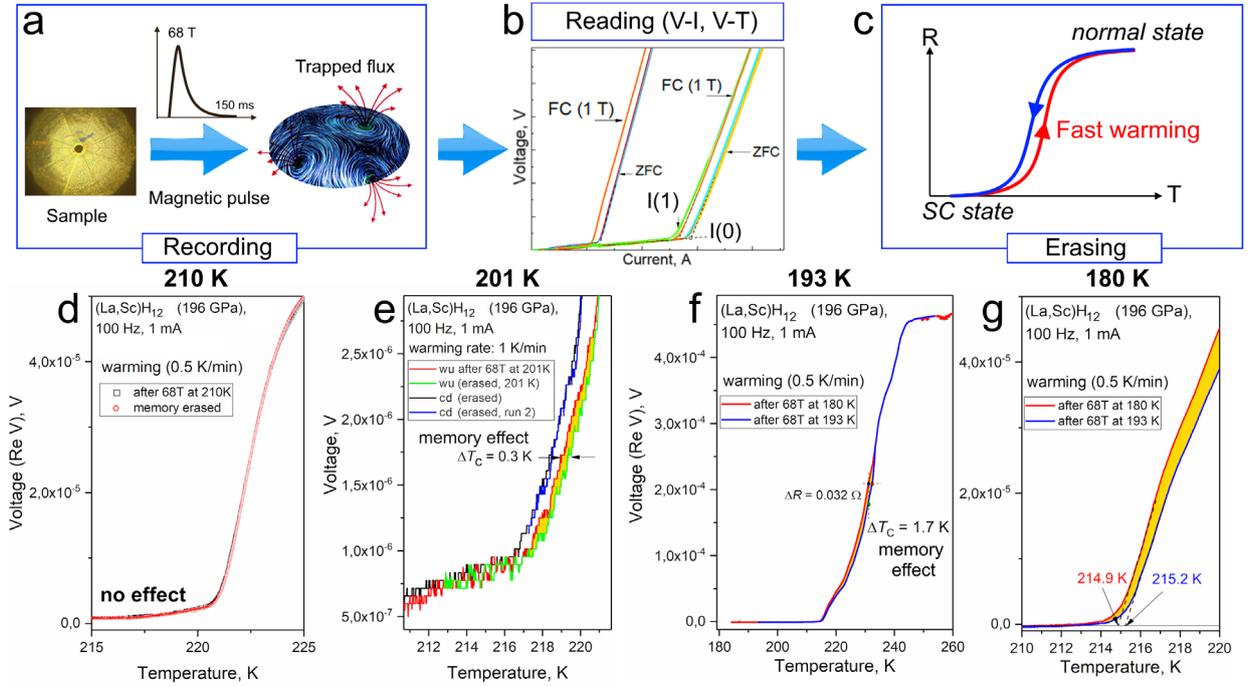

**Figure 4.** Scheme of recording and reading magnetic memory in superhydride samples and examples of its implementation in (La,Sc)H$_{12}$. (a) Memory recording using short magnetic field pulses in the superconducting state of the sample ($T < T_C$(offset)). (b) Reading of the memory, which can be realized using $V-I$ curves, $V-T$ curves or SQUID magnetometer [11]. (c) Memory erasure by means of a fast sample heating cycle to the normal state $T > T_C$(onset) with the destruction of all trapped vortices. After cooling (cd) to the SC state, the sample is again ready for recording. (d) Dependence of the voltage on the sample on the temperature ($V-T$) after a magnetic pulse of 68 Tesla at 210 K (black squares) and after complete "memory" erasure (red circles). Both $V-T$ curves practically coincide, the memory effect is absent. (e) $V-T$ curves after a magnetic pulse of 68 T at 201 K (red curve) and after complete erasure of the memory in the warming cycle (wu, green curve). The hysteresis zone is shown in yellow. Additionally, two $V-T$ curves obtained in the cooling (cd) cycle after erasure of the sample's memory are shown (black and blue curves). These curves practically coincide, which allows us to speak about a good reproducibility. (f) Comparison of two $V-T$ curves for the states with vortices trapped at 180 K (red curve) and 193 K (blue curve). The magnetic pulse at the lower temperature (180 K) leads to a stronger decrease in $T_C$. The hysteresis is 1.7 K, the resistance change is 0.032 Ω. (g) Zoom in of the hysteresis zone (highlighted in yellow) between the $V-T$ curves for the vortex states prepared at 180 K and 193 K.

We noticed that in the case of (La,Sc)H$_{12}$ a magnetic pulse of 68 T at 185 K leads to a significant decrease in the critical temperature $T_C$(offset) by 0.2 K when compared with the $T_C$ measured after the memory was erased (Figures 5a, d). Due to the thermal hysteresis between the warming and cooling cycles, in the following we compare only warming curves with a constant rate of 0.5 K/min.

Comparison of the $V-T$ curves after magnetic pulses at temperatures of 172 K, 178 K, 180 K and 193 K indicates a monotonic decrease in $T_C$ with decreasing the temperature when the magnetic pulse was made (Figures 5b, c, e, f). The lowest $T_C$(offset) is observed after the magnetic pulse at 172 K, and the highest for 193 K. Moreover, the largest difference (about 3-4 K) in $T_C$(offset) is observed between experiments with magnetic pulses at 210 K and 172 K, which makes the detection of the memory effect particularly simple.

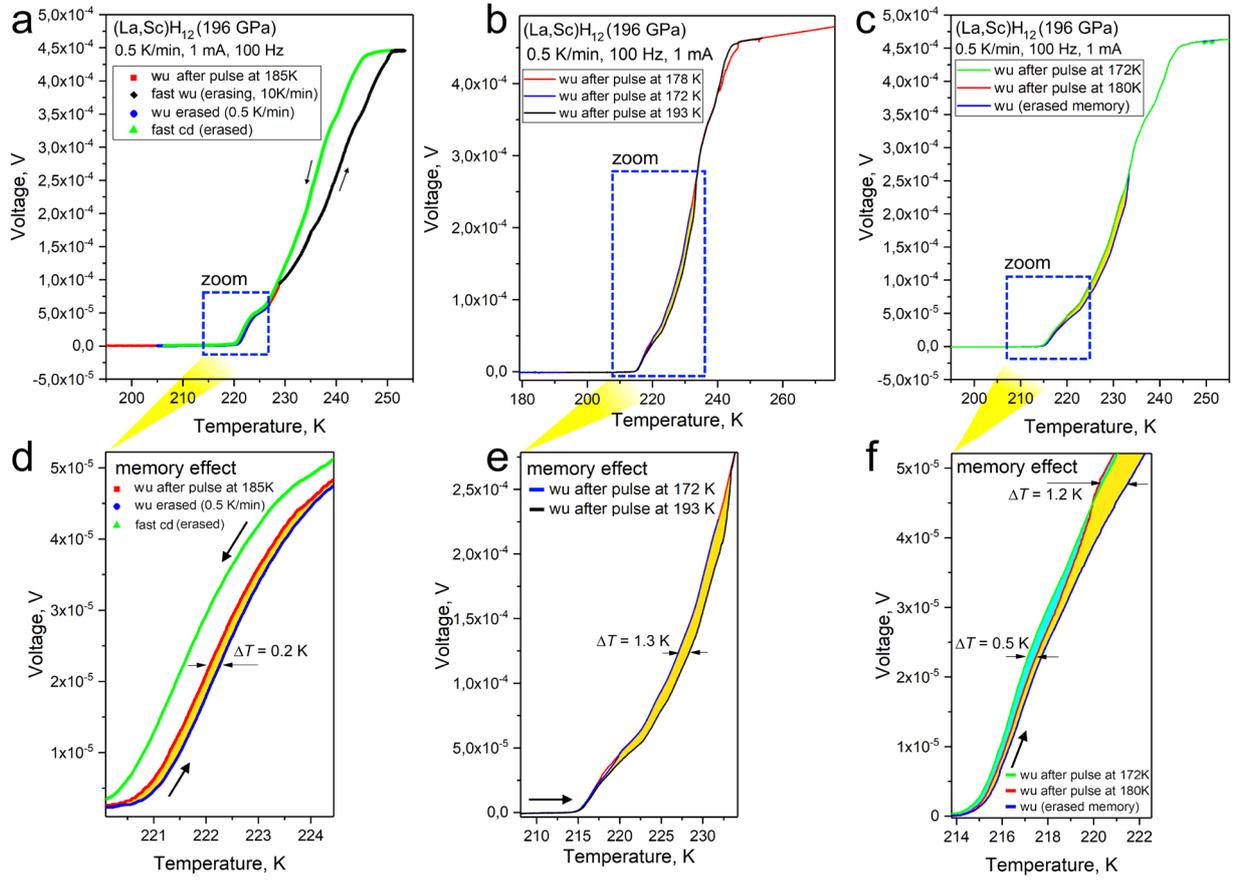

**Figure 5.** Hysteresis of the temperature dependence of voltage on the $(La,Sc)H_{12}$ sample at 196 GPa at different temperatures of the magnetic pulse (68 T). The hysteresis region is marked in yellow. (a, d) The magnetic field pulse was given at 185 K. Hysteresis (about 0.2 K) on the *V-T* curves in warming cycles (wu, rate is 0.5 K/min) is detected when comparing *V-T* measured immediately after the pulse and after the memory erasing cycle. (b, e) Comparison of *V-T* curves for a sample with different vortex states created by magnetic pulses at different temperatures: 172 K, 178 K and 193 K. The maximum hysteresis of the critical temperature was $\Delta T_C \approx 1.3$ K. (c, f) A similar comparison of V-T curves for vortex states created at 172 K (green curve) and 180 K (red curve), as well as for a sample with erased memory (blue curve).

As can be seen from Figures 3-5, the superconducting hydride sample "remembers" the magnetic pulse if the temperature at which the magnetic pulse was performed is low enough. Abrikosov vortices, trapped in the La-Nd and La-Sc hydrides, manifest themselves by decreasing the critical current and critical temperature of the sample, since they create an additional magnetic field in their vicinity. On the other hand, the applied excitation current is superimposed to the shielding currents induced by the trapped flux. As a result, the critical current density $J_C$ can be exceeded locally, and superconductivity is destroyed.

The observed hysteresis in the *V-I* and *V-T* characteristics disappears when the sample is heated to its normal state. This is the basis of the memory "erasing" cycle. Similar phenomena of magnetic flux trapping in superhydrides were detected earlier using a SQUID magnetometer in $H_3S$ and lanthanum polyhydrides $LaH_x$ [11], and also manifested in hysteresis when measuring the critical current in superconducting cuprates [1,2]. J. Hirsch in 2022 raised the question of the reason for the absence of hysteresis phenomena in transport measurements with superhydrides despite their granular nature [14]. Our work addresses this issue by providing evidence for the presence of pronounced hysteresis phenomena within simple transport measurements of critical currents and critical temperatures in lanthanum ternary polyhydrides. Moreover, together with studies on the detection of frozen magnetic flux using a SQUID magnetometer [11], and the discovery of a SQUID effect in hydrides [6], [7], this work is a proof of the possibility of implementing superconducting memory based on compressed hydrides at temperatures significantly higher than previously thought possible.

*3. Clockwise hysteresis loops in magnetoresistance of hydrides*

In the literature on granular cuprate superconductors, the majority of studies focus clockwise hysteresis loops in magnetoresistance (*R* vs *B* plot) and counterclockwise hysteresis loops in dependence of critical current on magnetic field ($I_C$ vs *B*) [1-5]. A comprehensive investigation of the magnetoresistance hysteresis in hydrides is beyond the scope of this study and will be addressed in future research. Nevertheless, the previously published steady-field magnetoresistance measurements of lanthanum-cerium (La, Ce)$H_{10+x}$ (where x = −1..+2) [7,24,25] and cerium $CeH_{9-10}$ [22,26] polyhydrides indicate that clockwise hysteresis is indeed observed for hydrides at different temperatures below $T_C$ (Figure 6). More detailed description of these experiments can be found in the aforementioned publications [7,22,24-26].

In granular superconductors [27], the expulsion of an external magnetic field *(B)* from the volume of SC granules leads to enhancement of magnetic field in the space between the granules and, consequently, to an increase in the magnetoresistance of the whole sample. A decrease in the magnetic field leads to a change in the sign of the magnetic moment of individual SC grains at a certain moment, and a weakening of the field between the granules. As a result, the magnetoresistance of the sample decreases and dependents on the sign of *dB/dt*.

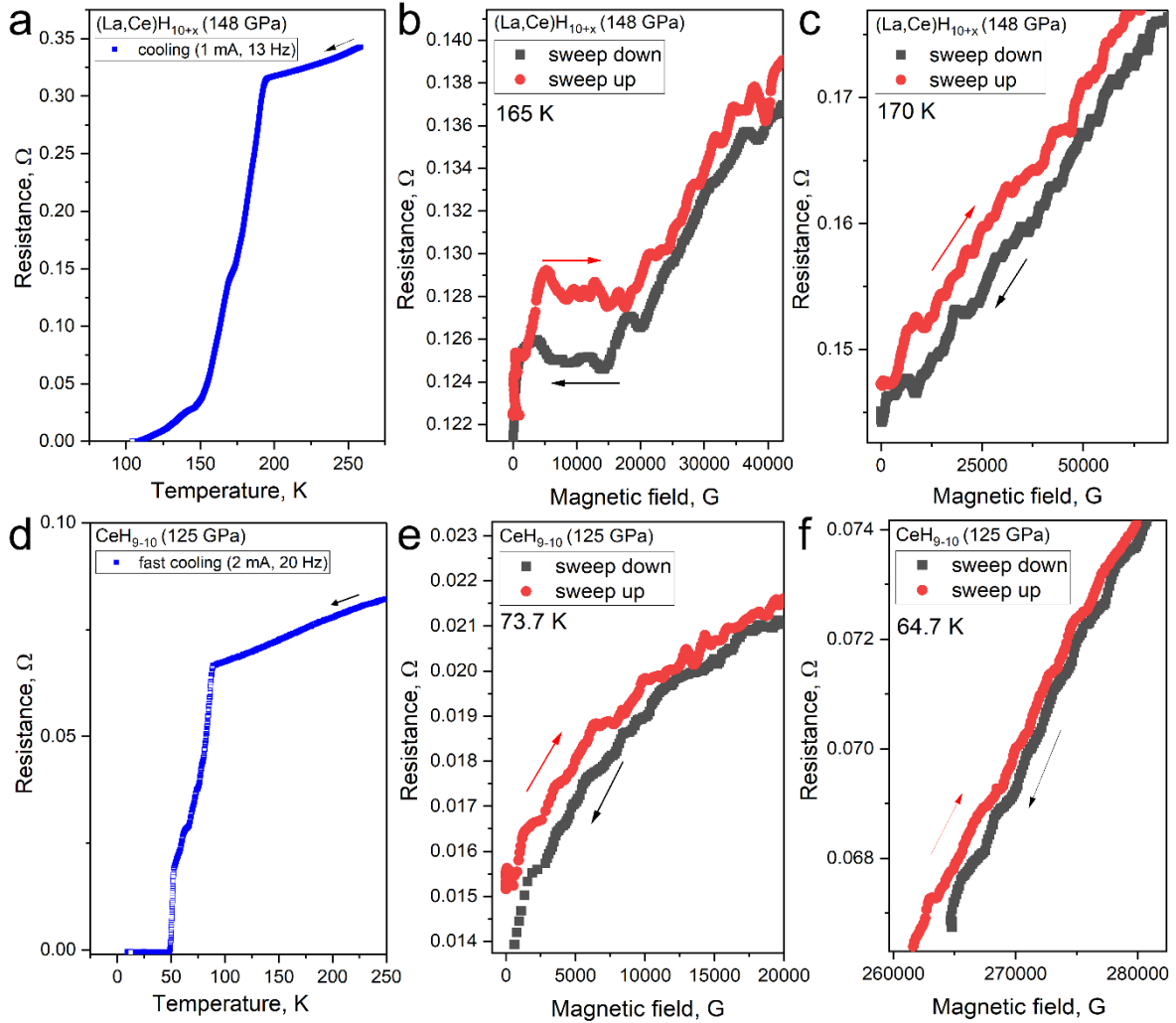

**Figure 6.** Clockwise magnetoresistance hysteresis in lanthanum-cerium (La, Ce)$H_{10+x}$ and cerium $CeH_{9-10}$ hydrides. (a) Dependence of the electrical resistance of (La, Ce)$H_{10+x}$ sample on the temperature in a cooling cycle at 148 GPa. Measurements were performed using AC current of 1 mA (RMS) with a frequency of 13 Hz. (b) Hysteresis loop of magnetoresistance observed in the La-Ce polyhydride at 165 K, and (c) at 170 K. (d) Dependence of the electrical resistance of $CeH_{9-10}$ sample on the temperature in a fast cooling (5-10 K/min)

cycle at 125 GPa. Measurements were performed using AC current of 2 mA (RMS) with a frequency of 20 Hz. (e) Hysteresis loop of magnetoresistance observed in the Ce polyhydride at 73.7 K, and (f) at 64.7 K.

The criticism of hydride superconductivity in Ref. [14] is based on trapped magnetic flux effects, which should be reflected in the properties of granular hydride superconductors. As we demonstrated in this paper, hysteresis effects associated with trapped magnetic flux, including clockwise hysteresis of magnetoresistance, are indeed observed in many polyhydrides. In a later work [28], J. Hirsch suggested the possibility of a metal-insulator transition observed, for instance, in $VO_2$ and $V_2O_3$, to explain the sharp drop in electrical resistance of polyhydrides. This includes multi-channel measurements using the van der Pauw scheme in La-Sc hydride $(La,Sc)H_{12}$ [6], sulfur deuteride $D_3S$, lanthanum $(LaH_{10})$ and sulfur hydrides $(H_3S)$ [29], where the electrical resistance drops by a factor of $10^3 - 10^5$ to below the noise level. Leaving aside the very possibility of such behavior of the van der Pauw scheme, and *R-T* measurements for $LaH_{10}$ in the Hall bar geometry [30], it is necessary to first of all note the significant hysteresis of the metal-insulator transitions, reaching tens of degrees [31,32]. However, resistive transitions in hydrides do not have thermal hysteresis [33,34], which calls into question J. Hirsch's argumentation.

## Conclusions

We discovered the magnetic memory effect in hydride superconductors and investigated it using two samples of lanthanum ternary superhydrides: $(La, Nd)H_{10}$ and $(La, Sc)H_{12}$ at 175 GPa and 196 GPa, respectively. The trapped magnetic flux reduces the critical temperature of the samples: $\Delta T_C$ is reaching 3-4 K in some cases. We have shown that the vortex state of the sample can be recorded using both a strong pulsed magnetic field (~ 68 T) and a sufficiently weak field of about 1 T or less. The vortex state of superhydride samples in DACs can be read-out using magnetic and electrical methods, in particular, by detecting the current-voltage characteristics, critical temperatures, and critical currents. The magnetic memory is erased by warming of the sample to a non-superconducting state. Polyhydride $(La,Sc)H_{12}$ gives an example of superconducting memory, written and stored at a record high temperature of 225-230 K, inaccessible to any other class of superconductors. Furthermore, we have demonstrated the existence of a clockwise hysteresis of magnetoresistance in the *R* vs *B* plots for Ce and La-Ce polyhydrides, which is a defining characteristic of granular superconductors.

## Acknowledgments

D. S. and D. Z. thank Beijing Natural Science Foundation (grant No. IS23017) and National Natural Science Foundation of China (NSFC, grant No. 12350410354) for financial support of this research. D. Z. also thanks China Postdoctoral Science Foundation (Certificate No. 2023M740204) and finical support from HPSTAR. D.S. and D.Z. are grateful for support from the EMFL-ISABEL project. The high-pressure experiments were supported by the Russian Science Foundation (Project No. 22-12-00163). A.V.S. acknowledges the financial support of RSF grant No. 24-72-10109. V.V.S. acknowledges the financial support from Shanghai Science and Technology Committee, China (No. 22JC1410300) and Shanghai Key Laboratory of Material Frontiers Research in Extreme Environments, China (No. 22dz2260800). A portion of this work was performed on the Steady High Magnetic Field Facilities (SHMFF), High Magnetic Field Laboratory, Chinese Academy of Science (CAS). This work was supported by HLD-HZDR, member of the European Magnetic Field Laboratory (EMFL).

## Contributions

I.A.T., D.V.S., D.Z., A.V.S., O.A.S., T.H., S.L. and K.P. performed the experiments. K.S.P. prepared the La–Sc and La-Nd alloys. I.A.T. prepared diamond anvil cells for experiments. A.V.S., O.A.S., and V.M.P. performed the magnetotransport experiments in magnetic fields below 16 T and participated in the data processing and discussions. T.H. and S.L. assisted in research in pulsed magnetic fields up to 68 T. D.V.S., V.M.P. and V.V.S. wrote the manuscript. All the authors discussed the results and offered useful inputs.

## Data availability

Authors declare that the main data supporting our findings of this study are contained within the paper and Supporting Information. All relevant raw data are available on GitHub: https://github.com/mark6871/Materials-Today-Physics-Memory-Effect.

## References

[1] D. A. Balaev, S. V. Semenov, D. M. Gokhfeld, J. Supercond. Nov. Magn. **36**, 1631 (2023).
[2] D. A. Balaev, D. M. Gokhfeld, A. A. Dubrovskiĭ, S. I. Popkov, K. A. Shaikhutdinov, M. I. Petrov, J. Exp. Theor. Phys. **105**, 1174 (2007).
[3] S. Sun, Y. Zhao, G. Pan, D. Yu, H. Zhang, Z. Chen, Y. Qian, W. Kuan, Q. Zhang, EPL **6**, 359 (1988).
[4] S. V. Semenov, A. D. Balaev, D. A. Balaev, Journal of Applied Physics **125** (2019).
[5] I. García-Fornaris, E. Govea-Alcaide, P. Muné, R. F. Jardim, Phys. stat. sol. (a) **204**, 805 (2007).
[6] D. V. Semenok and I. A. Troyan and D. Zhou and A. V. Sadakov and K. S. Pervakov and O. A. Sobolevskiy and A. G. Ivanova and M. Galasso and F. G. Alabarse and W. Chen, arXiv:2408.07477 (2024).
[7] D. V. Semenok, I. A. Troyan, D. Zhou, W. Chen, H.-k. Mao, V. V. Struzhkin, arXiv:2404.09199 (2024).
[8] A. V. Sadakov, V. A. Vlasenko, D. V. Semenok, D. Zhou, I. A. Troyan, A. S. Usoltsev, V. M. Pudalov, Phys. Rev. B **109**, 224515 (2024).
[9] A. V. Sadakov, V. A. Vlasenko, I. A. Troyan, O. A. Sobolevskiy, D. V. Semenok, D. Zhou, V. M. Pudalov, J. Phys. Chem. Lett. **14**, 6666 (2023).
[10] V. S. Minkov, S. L. Bud'ko, F. F. Balakirev, V. B. Prakapenka, S. Chariton, R. J. Husband, H. P. Liermann, M. I. Eremets, Nat. Commun. **13**, 3194 (2022).
[11] V. S. Minkov, V. Ksenofontov, S. L. Bud'ko, E. F. Talantsev, M. I. Eremets, Nat. Phys. **19**, 1293 (2023).
[12] E. Persky and A. V. Bjørlig and I. Feldman and A. Almoalem and E. Altman and E. Berg and I. Kimchi and J. Ruhman and A. Kanigel and B. Kalisky, Nature **607**, 692 (2022).
[13] R. Fermin, N. M. A. Scheinowitz, J. Aarts, K. Lahabi, Physical Review Research **4**, 033136 (2022).
[14] J. E. Hirsch, J. Supercond. Nov. Magn. **35**, 2731 (2022).
[15] D. V. Semenok, I. A. Troyan, A. V. Sadakov, D. Zhou, M. Galasso, A. G. Kvashnin, A. G. Ivanova, I. A. Kruglov, A. A. Bykov, K. Y. Terent'ev et al., Adv. Mater. **34**, 2204038 (2022).
[16] E. F. Talantsev, IEEE Transactions on Applied Superconductivity **34**, 1 (2024).
[17] A. D. Grockowiak, M. Ahart, T. Helm, W. A. Coniglio, R. Kumar, K. Glazyrin, G. Garbarino, Y. Meng, M. Oliff, V. Williams et al., Frontiers in Electronic Materials **2** (2022).
[18] F. Bloch, Z. Physik **59**, 208 (1930).
[19] F. J. Blatt, *Physics of Electronic Conduction in Solids* (McGraw-Hill, 1968).
[20] P. Allen, R. Dynes, Technical Report **7** TCM/4/1974 (1974).
[21] I. A. Troyan, D. V. Semenok, A. G. Ivanova, A. V. Sadakov, D. Zhou, A. G. Kvashnin, I. A. Kruglov, O. A. Sobolevskiy, M. V. Lyubutina, T. Helm et al., Adv. Sci. **2303622**, 1 (2023).

[22] D. Semenok, J. Guo, D. Zhou, W. Chen, T. Helm, A. Kvashnin, A. Sadakov, O. Sobolevsky, V. Pudalov, V. Struzhkin *et al.*, arXiv:2307.11742v2 (2023).

[23] J. Guo, G. Shutov, S. Chen, Y. Wang, D. Zhou, T. Cui, X. Huang, D. Semenok, Natl. Sci. Rev., nwae149 (2024).

[24] J. Bi and Y. Nakamoto and P. Zhang and K. Shimizu and B. Zou and H. Liu and M. Zhou and G. Liu and H. Wang and Y. Ma, Nat. Commun. **13**, 5952 (2022).

[25] W. Chen, X. Huang, D. V. Semenok, S. Chen, D. Zhou, K. Zhang, A. R. Oganov, T. Cui, Nat. Commun. **14**, 2660 (2023).

[26] W. Chen, D. V. Semenok, X. Huang, H. Shu, X. Li, D. Duan, T. Cui, A. R. Oganov, Phys. Rev. Lett. **127**, 117001 (2021).

[27] G. Deutscher, *New Superconductors: From Granular to High Tc* (2006).

[28] J. E. Hirsch, Journal of Superconductivity and Novel Magnetism (2024).

[29] F. F. Balakirev, V. S. Minkov, A. P. Drozdov, M. I. Eremets, Journal of Superconductivity and Novel Magnetism (2024).

[30] L. Sun, private communication (2024).

[31] J. A. Ramirez-Rincon, C. L. Gomez-Heredia, A. Corvisier, J. Ordonez-Miranda, T. Girardeau, F. Paumier, C. Champeaux, F. Dumas-Bouchiat, Y. Ezzahri, K. Joulain *et al.*, Journal of Applied Physics **124** (2018).

[32] Y. G. Liang, S. Lee, H. S. Yu, H. R. Zhang, Y. J. Liang, P. Y. Zavalij, X. Chen, R. D. James, L. A. Bendersky, A. V. Davydov *et al.*, Nature communications **11**, 3539 (2020).

[33] I. A. Troyan, D. V. Semenok, A. G. Ivanova, A. G. Kvashnin, D. Zhou, A. V. Sadakov, O. A. Sobolevsky, V. M. Pudalov, A. R. Oganov, Phys. Usp. **65**, 748 (2022).

[34] J. E. Hirsch, F. Marsiglio, arXiv:2101.07208 (2021).